\magnification1200

\rightline{KCL-MTH-05-13}
\rightline{hep-th/0511153}

\vskip .5cm
\centerline
{\bf    $E_{11}$, ten forms and supergravity }
\vskip 1cm
\centerline{Peter West}
\centerline{Department of Mathematics}
\centerline{King's College, London WC2R 2LS, UK}

\vskip .2cm
\noindent 
We extend the previously given non-linear realisation of $E_{11}$ for
the   decomposition appropriate to IIB supergravity to include the ten
forms that were known to be present in the adjoint representation. We find
precise agreement with the results  on ten forms found by closing the IIB
supersymmetry algebra. 
\vskip .5cm

\vfill
\eject
Long ago it was realised that the maximal supergravity theories in ten
and eleven dimensions  [1-4] when dimensionally  reduced on a torus 
 lead to maximal supergravity theories which possessed unexpected
symmetries.  In particular,  the eleven dimensional supergravity theory
when  dimensionally  reduced on a torus of
dimension $n$ possess an $E_n$ symmetry for
$n\le 8$ [5],  with some evidence [6] for an $E_9$ symmetry when reduced
to two dimensions and it has been conjectured [7] to have a $E_{10}$
symmetry in one dimension. 
 The scalar
fields which are created by the dimensional reduction process belong to
a coset, or non-linear realisation,  based on an  
$E_n$ algebra with the local sub-algebra being the Cartan involution
invariant sub-algebra. 
\par
In more recent years, it was realised [8] that the
entire bosonic sector of of eleven dimensional supergravity, including
gravity, could be formulated as a non-linear realisation [4]. In this
construction, the presence of gravity requires an $A_{10}$ algebra
together with   other 
 generators, which  transform as tensors under this $A_{10}$
algebra, and have non-trivial commutation relations amongst themselves
that are determined by the dynamics of the theory.  When formulated in
this way  it becomes apparent  that the eleven dimensional
supergravity theory may be part of a larger theory,  and assuming that
this is  a non-linearly realised Kac-Moody algebra,  one finds that it
must contain a    rank eleven algebra  called 
$E_{11}$ [9].  
A similar chain of argument applies to the bosonic sectors of the IIA and
IIB supergravity theories which are also thought to be part of  larger
theories that are  non-linear realisations of  $E_{11}$ [9,10]. 
\par
Similar ideas were subsequently taken up by the authors of reference [14]
who considered the idea that the eleven dimensional supergravity theory
is a non-linear realisation of the $E_{10}$ sub-algebra of $E_{11}$.
However, these authors proposed that space-time was in fact contained 
within $E_{10}$. A hybrid proposal based on $E_{11}$, 
but adopting similar ideas to the latter for space-time was also given
[15]. 
\par
We invite the reader to draw the Dynkin diagram of $E_{11}$ by drawing
ten nodes connected together by a single horizontal line. We label these
nodes from left to right  by the integers from one to ten and then add a
further node, labeled eleven, above node eight and attached by a single
vertical line. The latter node is sometimes called the exceptional node. 
We refer the reader to earlier works of the author for a  brief
review of Kac-Moody algebras useful for the considerations of this paper. 
\par
The eleven dimensional, IIA and IIB theories are thought to all have an
underlying 
$E_{11}$ symmetry which is non-linearly realised with the local
sub-algebra being the Cartan involution invariant sub-algebra. As a
result, in the non-linear realisation the group element contains positive
root and Cartan sub-algebra generators whose coefficients  turn out to be
the fields of the theory. The gravity sector  is
associated with a
$A_{D-1}$ type sub-algebra, where $D$ is the space-time
dimension of the theory, which arises  as a sub-Dynkin diagram  that 
contains node one and a set of continuously  connected nodes of the
$E_{11}$ Dynkin diagram. This set of nodes is referred to as the gravity
line. The eleven dimensional, IIA and IIB theories  are
distinguished by their different gravity sub-algebras, or gravity lines.
The eleven dimensional theory must possess an 
$A_{10}$ gravity algebra  and there is only one such algebra whose
gravity line contains all the nodes except node eleven. For this theory
it is useful to classify the $E_{11}$ algebra in terms of generators that
transform under this $A_{10}$ sub-algebra. 
\par
The IIA and IIB theories are ten dimensional and  to find 
these theories we seek an $A_9$ gravity sub-algebra 
and so we must  choose
the  gravity line to be  a 
 sub-Dynkin diagram that consist of nine nodes.  
Looking at the
$E_{11}$ Dynkin diagram there are only two ways to do this.  Starting
from the node labeled one  we must choose a $A_9$ sub-Dynkin diagram,
but once we get to  the
junction of the $E_{11}$ Dynkin diagram, situated at the node labeled 8,
 we can continue along the horizontal line 
with two further nodes taking only the
first node to belong to the $A_9$, or we can find the final $A_9$ node by
taking it to be the only node  in the other choice of direction at the
junction.  These two ways correspond to
the IIA and IIB theories respectively. Hence, in the IIA theory we take
the gravity line to be nodes labeled one to nine inclusive while for the
IIB theory the gravity line contains nodes one to eight and in addition
node eleven [9,10]. For these two theories it is useful to classify the
$E_{11}$ algebra in terms of their respective $A_9$ sub-algebras, but as
these are different embeddings in $E_{11}$ we find different field
contents. 
\par
While the number and type of generators is not known for any Kac-Moody
algebra one can find them at low levels. Every generator corresponds
to a root in the Kac-Moody algebra which can be written in terms of an
integer sum of the simple roots. By definition a Lorentzian Kac-Moody
algebra is one  which possess a Dynkin diagram which has  one 
  node whose deletion leads to  a Dynkin diagram that corresponds to
finite algebra together with possibly only one affine algebra [16].  For
the
$E_{11}$ Dynkin diagram we may delete node eleven to obtain an $A_{10}$
sub-algebra and so  $E_{11}$ is a Lorentzian Kac-Moody algebra. The
advantage of such algebra is that one can study its properties in terms
of the remaining sub-algebra, or algebras,  whose representations are well
known. In particular we may decompose the Lorentzian algebra, meaning its
adjoint representation,  into representations of the sub-algebra.  The
representations of the latter are determined by their highest weights.  
 A given
highest weight will appear in a particular root of the Lorentzian
algebra and the number of times the roots of the deleted nodes appear in
this root are  called the levels and can be used to label the
representations of the sub-algebra that appear in the decomposition
[14,17].  For example, deleting node eleven in the $E_{11}$ Dynkin
diagram we obtain an $A_{10}$ sub-algebra whose decomposition with
respect to which is appropriate to the eleven dimensional theory.
Carrying this out, one finds at low levels that the algebra  contains
the generators of
$A_{10}$, and then a three form and six form
generator as well as a generator with eight indices anti-symmetrised and
a further index. In the non-linear realisation these generators
correspond to gravity, the three form gauge field and its dual, and  
dual graviton respectively, which is the field content of the eleven
dimensional supergravity theory [9]. There are, of course,  an infinite
number of generators,  and so fields, at higher levels. 
\par
By deleting nodes nine and ten we decompose the
$E_{11}$ algebra with respect to an $A_9$ algebra that is the one
appropriate to the IIB theory. The representations are labeled by two
integers corresponding to the nodes deleted and are listed in the table
on page 27  of reference  [12]. As first noticed in reference [10] one
finds at low levels a set of generators that correspond precisely to the
field content of the IIB supergravity theory and their duals. Indeed, if
one includes the dual of gravity,  it is very striking how  this accounts
for the first nine entries of the table.  However, there are at higher
levels  an infinite number of other fields. Among
these one finds an additional eight form which, together with the earlier
ones    form an SU(1,1) triplet. 
One also finds 
some  ten
forms which form a doublet and quadruplet of SU(1,1) [12]. 
The triplet of eight forms  was first observed in
reference [18]. Although the ten form 
fields have no dynamics they couple to space-filling branes which are
dynamical. Their existence was first considered as a result of
 a string world-sheet analysis of D branes considerations [21]. 
Two ten forms were also observed in the context of IIB supergravity in 
reference [23], but it was shown in reference [24] that these could not
be a doublet of SU(1,1).  The eight and ten form fields fields have more
subsequently been seen from an entirely different view point. The authors
of reference [11] considered what eight and ten forms could be added to
the IIB theory  such that the supersymmetry algebra still closed. They
found precisely the  eight and ten forms predicted by
$E_{11}$. These authors also found, using the same calculation,  the gauge
transformations of all the gauge fields including the eight and ten form
fields and constructed some gauge invariant quantities.  
\par
In this paper we extend the calculation of reference [10] to include
one further eight form and the ten forms and compute the $E_{11}$
invariant Cartan forms constructed from the gauge fields.  We find that
these are in precise agreement with gauge invariant objects computed using
the closure of the supersymmetry algebra in   reference  [11]. 
\par
The Kac-Moody algebra $E_{10}$ considered in reference [14] does not
possess [20] the ten forms that occur in the $E_{11}$ theory and whose
presence has been confirmed in the IIB supergravity theory.  
\par

As explained above, the IIB theory emerges from the $E_{11}$  algebra by
taking the  decomposition with respect to a particular
$A_9$ algebra, hence forth denoted $\hat A_9$,   whose Dynkin diagram is
embedded in that of
$E_{11}$ by taking nodes labeled one to eight and  node eleven. The
latter is the so called exceptional node of the algebra. Carrying out the
non-linear realisation one finds that the 
$\hat A_9$ algebra is associated with the gravity fields of the IIB
theory and we denote its generators by $\hat K^a{}_b$.
The nodes not
included in the $\hat A_9$ sub-algebra, or gravity line, 
are the nodes labeled nine and ten of the
$E_{11}$ Dynkin diagram which are therefore the ones that must be deleted
to find the
$\hat A_9$ decomposition of the $E_{11}$ algebra.  The  $\hat A_9$
representations in the decomposition are then labeled by the levels
associated with these two nodes, that is the number of times these
two roots occur in the $E_{11}$ root that contains   the $\hat A_9$
highest weight of the representation under consideration. The
decomposition with this labeling is given in table on page 27  in
reference [12].  The 
$E_{11}$  algebra is generated by the Chevalley generators 
$H_a, E_a ,F_a, a=1,\ldots ,11$.
The
SU(1,1) invariance of the IIB theory is easy to see from the $E_{11}$
view point as it is just the
$A_1$ algebra associated with  node ten. As this is not directly
connected to the gravity line of the IIB theory it is an internal
symmetry . Thus the SU(1,1) is generated by the $H_{10},E_{10}$ and
$F_{10}$ generators. In fact, one can just delete node nine as then the
$E_{11}$  Dynkin diagram splits into two pieces corresponding to $\hat
A_9$ and
$A_1$ which classify the representations corresponding to the deletion of
this node. It is straightforward to collect the generators in the 
table of reference [12]  at a
given level corresponding to the root $\alpha_9$    into multiplets
of  $A_1$. 
\par
When constructing the $E_{11}$ non-realisation the  $E_{11}$ group
element contains the Cartan sub-algebra elements and the positive root
generators whose coefficients are the fields of the theory. However,
the description of the $E_{11}$ algebra from the mathematical view
point does not lead to the usual fields that appear in the supergravity
theories. The latter, that is the  physical fields, arise as coefficients
of linear combinations of the generators used in the mathematical
formulation of the $E_{11}$ algebra.   In particular,  the  Cartan
sub-algebra of
$E_{11}$ contains the generators 
$H_a, a=1,\ldots ,11$ when formulated in terms of its Chevalley basis.
Their relation to the generators associated to the fields of the IIB
theory, which are given  a hat,  is given by [10]
 $$
H_a=\hat K^a{}_a - \hat K^{a+1}{}_{a+1}, a=1,\ldots, 8,\
H_{9}=\hat K^{9}{}_{9}+\hat K^{10}{}_{10}+\hat R-{1\over
4}\sum_{a=1}^{10}\hat K^a{}_a, 
$$
$$
   H_{10}=-2\hat R_1,\ H_{11}=\hat K^{9}{}_{9}-\hat K^{10}{}_{10}
\eqno(1)
$$ 
The generator $\hat R_1$ will turn out to be associated with the dilaton
${\sigma}$ of the IIB theory in the non-linear realisation as
traditionally normalised. 
\par
The positive root Chevalley generators $E_a, a=1,\ldots , 11$ of $E_{11}$
are given by [10] 
$$
E_a=\hat K^a{}_{a+1}, a=1,\ldots 8,\  E_9=\hat R_1^{9 10},\
E_{10}=\hat R_2,\  E_{11}=\hat K^9{}_{10}.
\eqno(2)
$$
where the generators $\hat R_1^{a b}$
and 
$\hat R_2$ are associated with the NS-NS two form and the axion, $\hat
\chi$ of the IIB theory respectively. 
The last equation 
reflects the fact that the node labeled eleven is the last node in the
IIB gravity line, but is the exceptional node of the $E_{11}$ algebra. 
\par
The   $E_{11}$ algebra is just multiple commutators of the $E_a$, and
separately the $F_a$ generators, subject to the Serre relations. However,
it is more efficient  to construct the algebra using the list of
generators with the  $\hat A_9$ decomposition given in the  table on page
27   of reference [12] and then ensuring that the Jacobi identities are
satisfied. This was done
 for all the generators which in the
non-linear realisation are associated with  the fields of the IIB
supergravity theory and their duals in reference [10] and extended to
higher levels in reference [13]. This construction also included the
generator corresponding to the  the dual of the gravity field,  two 
eight form generators, which are duals of the scalar fields  and in
addition one of the ten form generators. Examining the table of reference
[12], we find that it  contains at low levels three eight forms which
make up a triplet as well as  a doublet and quadruplet of ten forms. 
 We now  extend this
construction of the algebra  to include the third of the eight form
generators and all the other ten form generators. It will be 
advantageous to do this in such a way that the  
$A_1$ character of the fields are manifest. Since the part of
the theory we wish to test concerns the gauge fields we will not
explicitly discuss the generators $\hat K^a{}_b$ of the $\hat A_9$ and  set
to zero the generator
$R^{a_1\ldots a_7,b}$, corresponding to the dual of gravity,  when it
appears on the right hand side of the  commutators. Considering   IIB 
table  of reference [12], and taking the last comment into account, we
introduce the positive root generators of 
$E_{11}$  not in the Cartan sub-algebra in the form 
$$
E_{10}, T^{a_1a_2}_\alpha,  T^{a_1\ldots a_4},  T^{a_1\ldots a_6}_\alpha, 
 T^{a_1\ldots a_8}_{\alpha\beta},  T^{a_1\ldots
a_{10}}_{\alpha\beta\gamma}, T^{a_1\ldots a_{10}}_\alpha, \ldots 
\eqno(3)
$$
The  $E_{11}$ algebra for these generators is given by 
$$
[T^{a_1a_2}_\alpha,
T^{a_3a_4}_\beta]=-\epsilon_{\alpha\beta} T^{a_1\ldots a_4},\ 
[T^{a_1a_2}_\alpha,T^{a_3\ldots a_6}]=4T^{a_1\ldots a_6}_\alpha,\ 
[T^{a_1a_2}_\alpha,T^{a_3\ldots a_8}_\beta]=
-T^{a_1\ldots a_8}_{\alpha\beta}
$$
$$
[T^{a_1a_2}_\alpha, T^{a_1\ldots a_8}_{\beta\gamma}]=
T^{a_1\ldots a_{10}}_{\alpha\beta\gamma},\ 
[ T^{a_1\ldots a_4}, T^{b_1\ldots b_4}]=0,\ [T^{a_1\ldots a_6}_\alpha,
T^{b_1\ldots b_4} ]=0
\eqno(4)$$
The SU(1,1) properties of the generators are given by 
$$
[E_{10}, T^{a_1\ldots a_p}_{1\ldots 11 2\ldots 2}]=i_1
T^{a_1\ldots a_p}_{1\ldots 12 2\ldots 2}
\eqno(5)$$
where $i_1$ is the number of one indices and the generator on the
right-hand side of the commutator has one more 2 index than that in the
commutator. We also have 
$$
[H_{10}, T^{a_1\ldots a_p}_{\alpha_1\ldots \alpha_r}]=
(2(\alpha_1+\ldots +\alpha_r)-3r) T^{a_1\ldots a_p}_{\alpha_1\ldots
\alpha_r}
\eqno(6)$$
In deriving  these equations we have used that 
$E_{10}$ acts as a lower operator  for the representations of
SU(1,1), taking $T^{a_1a_2}_1=R^{a_1a_2}_1$, that is $T^{910}_1=E_9$,
and    normalising
$T^{a_1a_2}_2$ such that 
$[E_{10}, T^{a_1a_2}_1]=T^{a_1a_2}_2$. Then using equations (5) and
(6) and  the Jacobi identities, and the 
 defining relations $[H_{10},E_{10}
]=2E_{10}$ and
$[H_{10},E_{9} ]=-E_{9}$ we find the above equations. 
\par
The relation to the generators used in references [10] and [13]
 is given by 
$$
T^{a_1a_2}_\alpha= R^{a_1a_2}_\alpha, \ T^{a_1\ldots a_4}= R^{a_1\ldots
a_4}_2,\ T^{a_1\ldots a_6}_\alpha=-\epsilon_{\alpha\beta}R^{a_1\ldots
a_6}_\beta, \ T^{a_1\ldots a_8}_{11}=R^{a_1\ldots a_8}_2,\ 
$$
$$
 T^{a_1\ldots a_8}_{12}=-{1\over 2}R^{a_1\ldots a_8}_1,\ 
T^{a_1\ldots a_8}_{22}=-S^{a_1\ldots a_8}_2,\ 
 T^{a_1\ldots a_{10}}_{111}=R^{a_1\ldots a_{10}}_2
\eqno(7)$$
from which one can verify that, in the absence of one of the eight form
generators and three of the ten form generators, the above commutators
agree with those of references [10] and [13].  Although it may appear
that at first sight the  generator
$T^{a_1\ldots a_{10}}_\alpha$ can appear on the right hand side of 
lower level commutators it turns out that this is forbidden by the Jacobi
identities??.  
\par
The non-linear realisation is by definition a theory which is invariant
under
$g\to g_0gh$ where
$g_0$ is a constant $E_{11}$ transformations and $H$ is an element of the
local sub-algebra which in this case is the Cartan involution invariant
sub-algebra. We may use the latter to gauge away all the negative root
terms in the expression for the group element $g$. As such to  
  construct the non-linear realisation we consider the $E_{11}$ group 
element given by 
$$
g=e^{{B^{\alpha}_{a_1\ldots a_{10}}\over 10!}
T_{\alpha}^{a_1\ldots a_{10}}}
e^{{B^{\alpha\beta\gamma}_{a_1\ldots a_{10}}\over 10!}
T_{\alpha\beta\gamma}^{a_1\ldots a_{10}}}
e^{{B^{\alpha\beta}_{a_1\ldots a_{8}}\over 8!}
T_{\alpha\beta}^{a_1\ldots a_{8}}}
e^{{B^{\alpha}_{a_1\ldots a_{6}}\over 6!}
T_{\alpha}^{a_1\ldots a_{6}}}
$$
$$.e^{{B_{a_1\ldots a_{4}}\over 4!}
T^{a_1\ldots a_{4}}}
e^{{B^{\alpha}_{a_1 a_{2}}\over 2!}
T_{\alpha}^{a_1 a_{2}}} g_{A_1}
\eqno(8)$$
where 
$$
g_{A_1}=e^{\chi E_{10}}e^{\phi H_{10}}
\eqno(9)$$
We have, as previously stated, omitted the gravity sector. 
The Cartan forms are invariant under $g_0$ transformations and being part
of the Lie algebra are  of the form 
$$
g^{-1}\partial_\mu g=g_{A_1}^{-1} 
({\tilde G ^{\alpha}_{\mu a_1\ldots a_{10}}\over
10!} T_{\alpha}^{a_1\ldots a_{10}}+
{\tilde G ^{\alpha\beta\gamma}_{\mu a_1\ldots a_{10}}\over 10!}
T_{\alpha\beta\gamma}^{a_1\ldots a_{10}}
+{\tilde G ^{\alpha\beta}_{\mu a_1\ldots a_{8}}\over 8!}
T_{\alpha\beta}^{a_1\ldots a_{8}}
+{\tilde G ^{\alpha}_{\mu a_1\ldots a_{6}}\over 6!}
T_{\alpha}^{a_1\ldots a_{6}}
$$
$$+{\tilde G _{\mu a_1\ldots a_{4}}\over 4!}
T^{a_1\ldots a_{4}}
+{\tilde G ^{\alpha}_{\mu a_1 a_{2}}\over 2!}
T_{\alpha}^{a_1 a_{2}}) g_{A_1}
\eqno(10)$$
\par
Using equation (4) it is  straightforward to find that 
$$
\tilde G ^{\alpha}_{\mu a_1 a_{2}}= \partial_\mu 
B^{\alpha}_{ a_1a_{2}},\ 
\tilde G _{\mu a_1 \ldots a_{4}}= \partial_\mu B _{ a_1\ldots 
a_{4}}+3\epsilon_{\alpha\beta} B^{\alpha}_{ a_1a_{2}}\partial_\mu
B^{\beta}_{ a_3a_{4}} ,\ 
$$
$$
\tilde G _{\mu a_1 \ldots a_{6}}^\alpha= \partial_\mu B _{ a_1\ldots 
a_{6}}^\alpha-6.5.2  B^{\alpha}_{a_1a_{2}}( \partial_\mu B _{ a_3\ldots 
a_{6}}+\epsilon_{\gamma\delta} B^{\gamma}_{ a_3a_{4}}\partial_\mu
B^{\delta}_{ a_5a_{6}}),\ 
$$
$$
\tilde G _{\mu a_1 \ldots a_{8}}^{\alpha\beta}= \partial_\mu B _{ a_1\ldots 
a_{6}}^{\alpha\beta}+7.4 B^{(\alpha}_{[a_1a_{2}}( \partial_\mu B _{
a_3\ldots  a_{8}]}^{\beta)}-6.5  B^{\beta)}_{a_3a_{4}} \partial_\mu B _{
a_5\ldots  a_{8}]}
-3.5 B^{\beta)}_{a_3a_{4}}\epsilon_{\gamma\delta} B^{\gamma}_{
a_5a_{6}}\partial_\mu B^{\delta}_{ a_7a_{8}]})
$$
$$
\tilde G ^{\alpha\beta \gamma}_{\mu a_1\ldots a_{10}}=\partial_\mu
B ^{\alpha\beta \gamma}_{ a_1\ldots a_{10}}
-9.5 B^{(\alpha}_{[a_1a_{2}}( \partial_\mu B _{
a_3\ldots  a_{10}}^{\beta\gamma)}+2.7  B^{\beta}_{a_3a_{4}} \partial_\mu B
_{ a_5\ldots  a_{10}]}^{\gamma)}
$$
$$
-8.7.5 B^{\beta}_{a_3a_{4}}B^{\gamma)}_{a_5a_{6}}\partial_\mu
B _{ a_7\ldots  a_{10}]}
-7.6.2 B^{\beta}_{a_3a_{4}}B^{\gamma)}_{a_5a_{6}}
\epsilon_{\epsilon\delta} B^{\epsilon}_{
a_7a_{8}}\partial_\mu B^{\delta}_{ a_9a_{10}]})
$$
$$
\tilde G ^{\alpha}_{ a_1\ldots a_{10}}=\partial_\mu B^{\alpha}_{
a_1\ldots a_{10}}
\eqno(11)$$
\par
We denote the result of carrying out the evaluation of the final SU(1,1)
$g_{A_1}$ factors  by  
$$
g^{-1}\partial_\mu g=
{ G ^{\alpha}_{\mu a_1\ldots a_{10}}\over
10!} T_{\alpha}^{a_1\ldots a_{10}}+
{ G ^{\alpha\beta\gamma}_{\mu a_1\ldots a_{10}}\over 10!}
T_{\alpha\beta\gamma}^{a_1\ldots a_{10}}
+{ G ^{\alpha\beta}_{\mu a_1\ldots a_{8}}\over 8!}
T_{\alpha\beta}^{a_1\ldots a_{8}}
+{ G ^{\alpha}_{\mu a_1\ldots a_{6}}\over 6!}
T_{\alpha}^{a_1\ldots a_{6}}
$$
$$+{ G _{\mu a_1\ldots a_{4}}\over 4!}
T^{a_1\ldots a_{4}}
+{ G ^{\alpha}_{\mu a_1 a_{2}}\over 2!}
T_{\alpha}^{a_1 a_{2}}+S^1_\mu \tilde E_{10}+S^2_\mu  H_{10}
\eqno(12)$$
  Using
equations (5) and (6) one finds that 
$$
 G ^{\alpha}_{\mu a_1 a_{2}}= \tilde G ^{\beta}_{\mu a_1 a_{2}}
U_\beta{}^\alpha,\ 
 G ^{\alpha}_{\mu a_1 \ldots a_{6}}=\tilde G ^{\beta}_{\mu a_1 \ldots
a_{6}} U_\beta{}^\alpha,\ 
G ^{\alpha}_{\mu a_1\ldots  a_{4}}= \tilde G ^{\alpha}_{\mu a_1 \ldots
a_{4}}
$$
$$
 G ^{\alpha\beta}_{\mu a_1 \ldots a_{8}}=\tilde G ^{\delta \epsilon}_{\mu
a_1
\ldots a_{6}} U_\delta{}^\alpha U_\epsilon{}^\beta,\ 
 G ^{\alpha\beta\gamma}_{\mu a_1 \ldots a_{10}}= \tilde G ^{\delta
\epsilon\tau}_{\mu a_1
\ldots a_{10}} U_\delta{}^\alpha U_\epsilon{}^\beta
U_\tau{}^\gamma{},\  G ^{\alpha}_{\mu a_1 \ldots a_{6}}= \tilde G
^{\beta}_{\mu a_1 \ldots a_{6}}U_\beta{}^\alpha
\eqno(13)$$
where 
$$
U=\left(\matrix{e^\phi& -\chi e^{-\phi}\cr
0&e^{-\phi}\cr}\right)
\eqno(14)$$
The last two
terms in equation (12) are just the standard vierbein and connection
on the SU(1,1)/U(1) coset
\par
The Cartan forms are inert under rigid $E_{11}$
transformations, but transform under the local sub-algebra. They do not
contain the curl of the gauge fields and so are not invariant under gauge
transformations. However, a  rigid  $E_{11}$ transformation for a
particular generator shifts the field corresponding to that generator
as well as giving  field dependent terms.  This transformation can be
thought  of as a particular gauge
transformation. For example under a rigid  $E_{11}$ transformation
corresponding to the generator
$T^{a_1a_2}_\alpha$ we find that 
$\delta B_{a_1a_2}^\alpha= a_{a_1a_2}^\alpha+\ldots $ where
$a_{a_1a_2}^\alpha$ is a constant. This is a gauge transformation  
$\delta B_{a_1a_2}^\alpha= 2\partial _{[a_1}\lambda_{a_2]}^\alpha+\ldots $
with gauge parameter
$\Lambda_a^\alpha= {1\over 2}a_{a b}^\alpha x^b$. 
The Cartan forms of equation
(11) are used to construct the equations of motion, but 
to find the field equations of IIB supergravity [10] one used only a
sub-set of all the Cartan forms and for the fields with completely
anti-symmetrised indices this was  the totally anti-symmetrised
Cartan forms, that is the field strengths given by 
$$
F_{a_1\ldots a_{p+1}}^{\alpha_1\ldots \alpha_r}=(p+1) G_{[a_1\ldots
a_{p+1}]}^{\alpha_1\ldots \alpha_r} 
\eqno(15)$$
The $\mu$ index is converted to a tangent index using the delta symbol as
we are taking the gravity sector to be trivial.   One  way
to view this enforced anti-symmetrisation is to consider demanding that
the theory also be invariant under the simultaneous non-linear
realisation of the conformal group. For gravity alone this does pick out
particular combinations of the $A_{D-1}$ Cartan forms and one finds it
leads   uniquely to Einstein's theory [19]. Thus although one started with
rigid transformations one ended up with local general coordinate
transformations. In fact,  the closure of translations and  
$A_{D-1}$ transformations  leads to general coordinate transformations. 
For gauge fields it is also true that the closure of rigid transformations
arising from a non-linear realisation and conformal transformations lead
to local symmetries, namely gauge transformations [9]. When the maximal
supergravity theories in ten  and eleven transformations were found using
the $E_{11}$ non-linear realisation it was also combined with the
conformal group [9,10]. Should one not carry out  this latter step then
one would find the correct equations of motion, but some constant would
have to be chosen appropriately.  The result of the closure of conformal
transformations and 
$E_{11}$ transformations is unexplored, but it does
convert all the $E_{11}$ rigid transformations  into local
transformations and so the above rigid transformations into  gauge
transformations. We should note that in finding the  equations of motion
of the maximal supergravities from the non-linearly $E_{11}$ in
references [9,10] one only required the local Lorentz part of the
local sub-algebra and it would  be very instructive to enforce
the rest of the local sub-algebra up to the level required. 
\par
In order to compare the invariant quantities that arise with those in
reference  [11] we must carry out a field redefinition. In particular,
carrying out the  field redefinitions 
$$
C^{\alpha}_{a_1 a_{2}}=B^{\alpha}_{a_1a_{2}},\ 
C_{a_1\ldots a_{4}}=B_{a_1\ldots a_{4}},\ 
C^{\alpha}_{a_1\ldots a_{6}}= B^{\alpha}_{a_1\ldots a_{6}}
-5.8 B^{\alpha}_{[a_1a_{2}}B_{a_3\ldots a_{6}]}$$
$$
C^{\alpha\beta}_{a_1\ldots a_{8}}= B^{\alpha\beta}_{a_1\ldots
a_{8}} +3.7 B^{(\alpha}_{[a_1a_{2}} C^{\beta)}_{a_1\ldots a_{6}]}
 +7.5.4.3 B^{(\alpha}_{[a_1a_{2}}B^{\beta )}_{a_3a_{4}}B_{a_5\ldots
a_{8}]},\ 
$$
$$
C^{\alpha\beta\gamma}_{a_1\ldots a_{10}}=
B^{\alpha\beta\gamma}_{a_1\ldots a_{10}}-9.4 B^{(\alpha}_{[a_1a_{2}}
 C^{\beta\gamma )}_{a_3\ldots a_{10}]}
+9.7.5.3 B^{(\alpha}_{[a_1a_{2}}B^{\beta}_{a_3a_{4}}
C^{\gamma)}_{a_5\ldots a_{10}]}
$$
$$
+16.9.7.5  B^{(\alpha}_{[a_1a_{2}}B^{\beta}_{a_3a_{4}}
B^{\gamma )}_{a_5a_{6}}B_{a_7\ldots a_{10}]}
,\ 
C^{\alpha}_{a_1\ldots a_{10}}=B^{\alpha}_{a_1\ldots a_{10}}
\eqno(16)$$
the Cartan forms become   
$$
\tilde G ^{\alpha}_{\mu a_1 a_{2}}= \partial_\mu 
C^{\alpha}_{ a_1a_{2}},\ 
\tilde G _{\mu a_1 \ldots a_{4}}= \partial_\mu C _{ a_1\ldots 
a_{4}}+3\epsilon_{\alpha\beta} C^{\alpha}_{ [a_1a_{2}}
\tilde G^{\beta}_{\mu a_3a_{4}]} ,\ 
$$
$$
\tilde G _{\mu a_1 \ldots a_{6}}^\alpha= \partial_\mu C _{ a_1\ldots 
a_{6}}^\alpha
-5.4 C^{\alpha}_{[a_1a_{2}} \tilde G _{\mu a_3\ldots  a_{6}]}
+8.5 \tilde G^{\alpha}_{\mu [a_1a_{2}} C _{a_3\ldots  a_{6}]}, 
$$
$$
\tilde G _{\mu a_1 \ldots a_{8}}^{\alpha\beta}= \partial_\mu C _{
a_1\ldots  a_{6}}^{\alpha\beta}
+7  B^{(\alpha}_{[a_1a_{2}} \tilde G _{\mu a_3 \ldots a_{8}]}^{\beta)}
-7.3  \tilde G^{(\alpha}_{\mu [a_1a_{2}} C _{ a_3 \ldots
a_{8}]}^{\beta)}
$$
$$
\tilde G ^{\alpha\beta \gamma}_{\mu a_1\ldots a_{10}}=\partial_\mu
C ^{\alpha\beta \gamma}_{ a_1\ldots a_{10}}
-9 C^{(\alpha}_{[a_1a_{2}} \tilde G _{\mu a_3\ldots 
a_{10}]}^{\beta\gamma)} +9.4 \tilde G ^{(\alpha}_{\mu a_1a_{2}} C_{
a_3\ldots  a_{10}]}^{\beta\gamma)}
$$
$$
\tilde G ^{\alpha}_{ a_1\ldots a_{10}}=\partial_\mu C^{\alpha}_{\mu
a_1\ldots a_{10}}
\eqno(17)$$
\par
The field redefinitions  of equation (16) contain all possible terms 
and the coefficients are fixed uniquely by 
requiring that the resulting Cartan forms 
 can be expressed in terms of the  field with $p$ anti-symmetrised
indices, the  field with $p-2$ anti-symmetrised
indices,  $B^{(\alpha}_{[a_1a_{2}}$ and $\tilde G ^{\alpha}_{\mu a_1
a_{2}}$. That this can be done is  non-trivial as there are fewer 
coefficients  in the field redefinition of equation (16)
than  the number of terms required to be eliminated to bring the Cartan
forms into the above form. The simplest way to see this is to change the
coefficient of the third term in $\tilde G _{\mu a_1 \ldots
a_{8}}^{\alpha\beta}$ in equation (11) from $-6.5$ to an arbitrary
number and then carry out the field redefinition to bring it to the
required form; one finds that this is not possible unless the coefficient
is $-6.5$. Similar restrictions hold for the ten form. 
 Substituting the expressions of equation
(17) into the field strengths of equation (15)  we can compare the
results with the field strengths of equation (5.18-23) of reference [11]. 
As we have just  noted to bring the Cartan forms into the required
form of equation (17) is already a non-trivial check. 
While some terms are not directly comparable due to possible field
rescaling the ratio between the last two terms in 
$\tilde G _{\mu a_1 \ldots a_{6}}^\alpha$, $\tilde G _{\mu a_1 \ldots
a_{8}}^{\alpha\beta}$ and $\tilde G ^{\alpha\beta \gamma}_{\mu a_1\ldots
a_{10}}$ are independent of such transformations. We find that they are
precisely those given by the  $E_{11}$ calculation carried out in this
paper. 
The ratios associated with the six and eight forms were 
already contained in reference [10], but their uniqueness was not
stressed. 
\par
 It may seem that the ten form comparison with  the two
reference [11] is not legitimate  as the field strength in the  reference
[11] has eleven indices and so each term vanishes identically. However,
as explained  in that paper the meaning of the field strength for these
authors  is that it invariant under the gauge transformation of the ten
forms in any dimension, hence the unambiguous ratio is between  the
coefficients in the gauge transformation of the ten form in equation
(5.17) of reference [11].  It is straightforward to verify that the ten
form Cartan form 
$\tilde G ^{\alpha\beta \gamma}_{\mu a_1\ldots a_{10}}$ of equation
(17) is invariant under the rigid transformation 
$$
\delta C ^{\alpha\beta \gamma}_{ a_1\ldots a_{10}}=
a ^{\alpha\beta \gamma}_{ a_1\ldots a_{10}}
-9.4 C_{[a_1a_2}^{(\alpha} a ^{\beta \gamma )}_{ a_3\ldots a_{10}]}
+9C_{[a_1\ldots a_8}^{(\alpha\beta} a ^{ \gamma )}_{ a_9
a_{10}]}+O(C^2)
\eqno(18)$$
One could have derived this transformation by carrying  out an appropriate
rigid $g_0$ transformation on the group element of equation (8)
followed by   the field redefinition of equation (16). As explained
above we can convert this rigid transformation  to a gauge transformation
by taking 
$a ^{\alpha_1\ldots  \alpha_r}_{ a_1\ldots a_{p}}=p\partial_{[a_1}
\Lambda_{a_2\ldots a_p]}^{\alpha_1\ldots  \alpha_r}$. Carrying out this
last step and  and then redefining the gauge parameter so as to bring it
to the form given in reference [11] we find that 
$$
\delta C ^{\alpha\beta \gamma}_{ a_1\ldots a_{10}}=
\partial_{[a_1}\Lambda^{\alpha\beta \gamma}_{ a_2\ldots a_{10}]}
-2F_{[a_1\ldots a_9}^{(\alpha\beta} \Lambda ^{ \gamma )}_{
a_{10}]}
+8.4.3  F_{[a_1a_2a_3}^{(\alpha} \Lambda ^{\beta \gamma )}_{
a_4\ldots a_{10}]}  +O(C^2)
\eqno(19)$$
Comparing with equation (5.17) of reference [11] we find that the ratio
in question between the last two terms is the same. Clearly, we could
have carried out this comparison the other way round by converting the
gauge transformation to the required rigid transformation. This throws
light on the observation in reference [11] that the ten form field
strength is invariant in any dimension, it is not so much to do with a
symmetry that can be lifted in dimension, but more to do with the fact
that the Cartan forms for the ten form, which are non-vanishing,  are
invariant under rigid
$E_{11}$ transformations. 
\par
The Cartan forms are inert under rigid $E_{11}$ transformations, but
transform under the local sub-algebra as $g^{-1}\partial_\mu g\to 
h^{-1}g^{-1}\partial_\mu g h+h ^{-1}\partial_\mu  h$. To form an object
that transforms covariantly we introduce the operation $I(A)=I_c(-A)$
where $I_c$ is the action of the Cartan involution. It acts on group
elements as $I(k)=I_c (k^{-1}$ and $I(g_1g_2)= I(g_1)I(g_2)$. The
Chevalley generators behave under the Cartan involution as
$I_c(E_a)=-F_a$ and  
$I_c(H_a)=-H_a$. Since the local sub-algebra is by definition invariant
under the Cartan involution it follows that
$I(h)=h^{-1}$. As a result, the quantity $U_\mu=g^{-1}\partial_\mu
g+I(g^{-1}\partial_\mu g)$ transforms are $U_\mu\to
h^{-1}U_\mu h$ while  $w_\mu={1\over 2}(g^{-1}\partial_\mu
g-I(g^{-1}\partial_\mu g))$ behaves like a connection 
$w_\mu\to h^{-1}w_\mu h+ h^{-1}\partial_\mu h$  
The equations of
motion are to be built from
$U_\mu$ and $w_\mu$ so as to ensure their invariance under rigid
transformations.  As we will see below,  we will be interested in
first order equations, however, we note that
$\partial_\mu U_\nu+[w_\mu, U_\nu] $ is second order in derivatives, but
transforms covariantly. Looking at equation (12) we see that 
$U_\mu=S^1_\mu(E+F)+2S^2_\mu H_{10}+G_{\mu a_1a_2}^{\alpha}
(T^{a_1a_2}_{\alpha}-I_c(T^{a_1a_2}_{\alpha}))+ \ldots $ where
$+\ldots $ are higher level  generators. 
\par
At the lowest level the local sub-algebra contains the Lorentz algebra and
the U(1) sub-algebra of the SU(1,1) algebra. The latter U(1) has the
generator $E_{10}-F_{10}$ and  it transforms the Cartan forms as 
$\delta U_\mu=-a[E_{10}-F_{10},U_\mu]$ where $a$ is the local parameter. 
Introducing $S^\pm=S^1\mp {2i}S^2$ we find it transforms as
$\delta S^{\pm\pm} =\pm 2ia S^{\pm\pm} $. The transformations of the other
fields are most easily displayed by introducing the analogue to light-cone
coordinates in the SU(1,1) index space; 
$$
T^{a_1a_2}_{\pm}={1\over 2}(T_1^{a_1a_2}\mp i T_2^{a_1a_2}),\ 
T^{a_1\ldots a_6}_{\pm}={1\over 2}(T^{a_1\ldots a_6}_1\mp i T^{a_1\ldots
a_6}_2),\ 
$$
$$
T^{a_1\ldots a_8}_{\pm\pm}={1\over 4}(
T^{a_1\ldots a_8}_{11}- T^{a_1\ldots a_8}_{22}\mp 2iT^{a_1\ldots
a_8}_{12}),\ 
 T^{a_1\ldots a_8}_{+-}={1\over 4}(
T^{a_1\ldots a_8}_{11}+ T^{a_1\ldots a_8}_{22})
\eqno(20)$$
Their U(1) commutators are given by 
$$
[E_{10}-F_{10},T^{a_1a_2}_{\pm}]=\pm i T^{a_1a_2}_{\pm},\ 
[E_{10}-F_{10}, T^{a_1\ldots a_6}_{\pm}]=\pm i T^{a_1\ldots a_6}_{\pm},
$$
$$ 
[E_{10}-F_{10},T^{a_1\ldots a_8}_{\pm\pm}]=\pm 2i T^{a_1\ldots
a_8}_{\pm\pm},\  [E_{10}-F_{10},T^{a_1\ldots a_8}_{+-}]=0
\eqno(21)$$
Introducing the analogous basis for the derivatives of the fields
that appear in the Cartan forms 
$$
G_{\mu a_1a_2}^{\pm}={1\over 2}(G_{\mu a_1a_2}^1 \mp i
G_{\mu a_1a_2}^2),\  G_{\mu a_1\ldots a_6}^{\pm}={1\over 2}(G_{\mu
a_1\ldots a_6}^1\mp i G_{\mu a_1\ldots a_6}^2),\ 
$$
$$
G_{\mu a_1\ldots a_8}^{\pm\pm}={1\over 4}(
+G_{\mu a_1\ldots a_8}^{11}- G_{\mu a_1\ldots a_8}^{22}\mp 2iG_{\mu
a_1\ldots a_8}^{12}),\ 
 G_{\mu a_1\ldots a_8}^{+-}={1\over 4}(
G_{\mu a_1\ldots a_8}^{11}+ G_{\mu a_1\ldots a_8}^{22})
\eqno(22)$$
and defining the U(1) charge by $\delta \bullet= 
[E_{10}-F_{10},\bullet]-q\bullet$ where $\bullet $ is any of the above we
find,  using equation (20), that the expressions in equation (22) have the
U(1) weights $\pm 1,\pm 1,\pm 2$ and $0$ respectively. 
\par
If we assume that the equations of motion for the gauge fields are first
order in space-time derivatives they are then uniquely specified by
demanding rigid $E_{11}$ invariance, which is guaranteed by using the
Cartan forms $U$,  and invariance under the  Lorentz and U(1) part of the
local sub-algebra;  
 $$
 F ^{\pm}_{ a_1 a_2 a_3}= {1\over 7!} \epsilon_{ a_1 a_2 a_3}{}^{
b_1\ldots b_{7}} F ^{\pm}_{ b_1\ldots b_{7}},\ 
S ^{\pm\pm}_{ a}= {1\over 2.9!} \epsilon_{ a}{}^{
b_1\ldots b_{9}} F ^{\pm\pm}_{ b_1\ldots b_{9}},\ 
F ^{+-}_{ b_1\ldots b_{9}}=0
\eqno(23)$$
These are the same equations are found in 
 reference [12], except for the last equation, which constrains two of the
three rank nine field strength to be equal. This last equation was given
in reference [11]. It would be of interest to test the invariance of
these equations at higher levels. 
\par

It was know [10] that 
the $E_{11}$ non-linear realisation  with only two of the three eight
branes and all lower forms lead to the bosonic equations of
motion of IIB supergravity. 
In this paper we have carried out the $E_{11}$ non-linear realisation
appropriate for the IIB theory including all the thee eight and ten forms
and  we have compared our results for the ten forms with those of
reference [11]   and found perfect agreement, including numerical
coefficients. While the calculation given in this paper is just an
exercise in $E_{11}$ algebra, the results  of reference [11] follow from
the closure of the IIB supersymmetry algebra. There would seem to be no
overlap between these two methods and so one can regard the results of
this paper as a rather non-trivial check on the
$E_{11}$ conjecture. 
\par
We could
have carried out the comparison with reference [11] in another way namely
by simply computing the algebra of rigid
$E_{11}$ transformations converted these to gauge transformations and
after a field redefinition carried out a comparison with the gauge
transformation of reference [11]. However, the results will be the same
as comparing the covariant objects as we have done in this paper. 
\par
The ten form does not possess a gauge invariant field strength so one
might expect that it has trivial dynamics, nonetheless it does couple, in
the supersymmetric Born-Infeld   action,  to a space-filling brane. This
does have  propagating field and as a result the ten form
and its transformation properties do have consequences for the dynamics
of the theory. In this context we note that it has been conjectured that
the brane dynamics should also be $E_{11}$ invariant [22]. 
\par
In the table on page 27 of reference [12] the lowest level ten form is at
the eighteenth (eleventh in terms of SU(1,1) multiplets) entry and has
level (4,5) and so one has now confirmed the presence of  fields in the
adjoint representation of
$E_{11}$ which are relatively far down the table. It is also interesting
to note that the $E_{11}$ root associated with some of the ten forms has
length squared $-2$ instead of the usual $2$ that occur in finite
dimensional semi-simple Lie algebras and the zeros that occur in 
affine algebra. A glance at the
table shows that it also possess in the vicinity of the ten forms a
SU(1,1) doublet of generators with the indices
$R^{a_1\ldots a_9,b}$ and also a doublet of generators of the form 
 $R^{a_1\ldots a_8,bc}$. It would be interesting to see if these can also
be seen from the viewpoint of the IIB supersymmetry algebra. One could
even wonder if one could find the dual gravity  field in such a
calculation. 
\par
As we have noted, at low levels the Borel sub-algebra generators in the
decomposition of
$E_{11}$ to the IIB theory are in a one to one correspondence with the
fields of IIB supergravity. As the latter can be assigned to either the
NS-NS or R-R sector of the IIB string theory, we can assign the
low level generators of $E_{11}$ to either the NS-NS or R-R sector. It was
observed in reference [13] that one can extend this classification to all 
the generators of $ E_{11}$ by taking the rule that the commutators admit
a  grading with the  R-R generators  being assigned as odd and NS-NS
generators as even. Looking at the table on page 27 of reference [12] one
see that a generator is even (odd), i.e. in the NS-NS (R-R) sector, if its
associated root has an even (odd) number of $\alpha_{10}$'s in its
decomposition into simple roots.  Put another way a generator with root
$\alpha$ is in the R-R (NS-NS) sector if $\alpha.\Lambda_{10}$ is odd
(even) where $\Lambda_{10}$ is the fundamental root associated with node
ten. As the roots add in any commutator this ensures the required graded
structure. We note that $\alpha.\Lambda_{10}$ is just the level $n_{10}$. 
Given this rule it is easy to assign the  ten
forms to either the generalised NS-NS or R-R sector. Looking at the
$E_{11}$ decomposed to the $\hat A_9$ sub-algebra appropriate to the IIA
theory given in the table on page 26 of reference [12] we find that a
similar assignment is allowed and that the NS-NS sector has an even level
corresponding to node ten and the R-R sector an odd level. 
\par
The eleven dimensional, IIA and IIB theories all are non-linear
realisations of $E_{11}$, but as there is only one $E_{11}$ with a
standard Chevalley presentation we can made a one to one correspondence
between the three theories [13]. Looking at the table of reference [12]
we see that all the ten forms in the IIB theory arise from the eleven
dimensional theory at level four, which is one level above that for the
dual graviton at level three and that below  level three one only has the
generators corresponding to the fields of eleven dimensional
supergravity.  In the IIB table we see that the ten forms have the 
$E_{11}$ roots 
$(1,2,3,4,5,6,7,8,5,a,4)$ with $a=1,2,3,4$.  That for $a=2$ has
multiplicity two and these are   easy to find in the  IIA table of
reference [12] as the two  ten forms 
 in  that table at low level  have a root of length squared $-2$,
also with multiplicity two and  precisely  the same $E_{11}$ root. That
these ten forms are related by T-duality is known to the authors of
reference [23].  For
$a=3$ which also has multiplicity two and length squared $-2$ we find the
same root lower down the table,  it corresponds to a IIA generator $\tilde
S^{10}$ that is the highest $\tilde A_9$ states of $\tilde S^a, a=1,\ldots
, 10$. The
$a=4$ root also appears in the IIA table and we find it is the highest
weight component of the generator $\tilde R^(ab)$. To find the last
member of the IIB quadruplet we use the fact that
$E_{10}$ and
$F_{10}$ raise and lower respectively in the same SU(1,1) multiplet. In
terms of IIA generators we have that $E_{10}=\tilde R^{10}$ where $\tilde 
R^{a}$ corresponds to the rank one gauge field in the IIA supergravity
theory. Acting with $F_{10}=\tilde R_{10}$, the latter being the
corresponding negative root,   on the
$a=2$ generator we will find the $a=1$ generator. This corresponds to the
commutator 
$[\tilde R_{10}, \tilde R^{1\ldots 10}]$ whose result is a generator with
nine indices $\tilde R^{1\ldots 9}$. However, this is not a highest
$\tilde A_9$ representation and so will not occur in the table. The
highest weight  is $R^{2\ldots 9}$ which is obtained by acting with
$\tilde K^2{}_1+\ldots +\tilde K^{10}{}_9$ which implies we must add the
root
$-(\alpha_1+\ldots +\alpha_9)$. As a result, we find a nine form whose
highest weight occurs in $E_{11}$ as the root $(0,1,2,3,4,5,6,7,4,1,4)$,
it is just the IIA nine forms $\tilde R^{a_1\ldots a_9}$ which is
associated with the  massive IIA theory. We note that the doublet of ten
forms in IIB  have the same roots as the $a=2,3$ members of the
quadruplet and so are also correspond to the ten form and $\tilde S^{10}$
in the IIA theory. 
IIA ten forms

\medskip 
{\bf Acknowledgments}
I wish to thank Fabio Riccioni for useful discussions.    This research
was supported by a PPARC senior fellowship PPA/Y/S/2002/001/44 and  in
part by the PPARC grants  PPA/G/O/2000/00451 and
the EU Marie Curie, research training network grant HPRN-CT-2000-00122. 
\medskip
{\bf References}
\medskip
\item{[1]} E. Cremmer, B. Julia and J. Scherk, 
{\it  Supergravity theory in eleven dimensions},  Phys. Lett. {\bf 76B}
(1978) 409.
 \item{[2]} I.C.G. Campbell and  P. West, {\it N=2 d=10 nonchiral
supergravity
     and its spontaneous compactifications}, Nucl. Phys. {\bf B243} (1984),
     112; M. Huq, M. Namanzie, {\it Kaluza-Klein supergravity in ten
     dimensions}, Class. Quant. Grav. {\bf 2} (1985); F. Giani, M. Pernici,
     {\it N=2 supergravity in ten dimensions}, Phys. Rev. {\bf D30} (1984),
     325
\item{[3]} J. Schwarz and  P. West {\it Symmetries and Transformations of
     chiral N=2, D=10 supergravity}, Phys. Lett. {\bf B126} (1983), 301.
\item{[4]}
     J. Schwarz, {\it Covariant field equations of chiral N=2 D=10
     supergravity}, Nucl. Phys. {\bf B226} (1983), 269; P. Howe and  P.
West,    {\it The complete N=2, d=10 supergravity}, Nucl. Phys. {\bf B238}
     (1984) 181. 
\item{[5]} E. Cremmer and B. Julia, {\sl The $N=8$ supergravity
theory. I. The Lagrangian}, Phys. Lett. {\bf B 80} (1978) 48; 
 N. Marcus
and J. Schwarz, {\it Three-dimensional supergravity theories}, Nucl.
Phys. {\bf B228} (1983) 301;  
 B.\ Julia, {\it Group disintegrations}, in
{\it   Superspace and
Supergravity}, p. 331,  eds. S. W. Hawking  and M.  Ro\v{c}ek,  
Cambridge University
Press (1981).
 \item {[6]} H. Nicolai, {\it The integrability of N=16  
supergravity},
  Phys. Lett. {\bf 194B} (1987) 402; {\it On M-Theory},  
hep-th/9801090.
\item {[7]} B. Julia, in {\it Vertex Operators in
Mathematics   and Physics},
Publications of the Mathematical Sciences Research Institute no 3,  
Springer
Verlag (1984); B. Julia in {\it Superspace and Supergravity} edited by  
S.W.
Hawking and M. Rocek, Cambridge University Press (1981).
\item{[8]} P.~C. West, {\sl Hidden superconformal symmetry in {M}
    theory },  JHEP {\bf 08} (2000) 007, {\tt hep-th/0005270}
\item{[9]} P. West, {\sl $E_{11}$ and M Theory}, Class. Quant.
Grav. {\bf 18 } (2001) 4443,  hep-th/010408. 
\item{[10]} I. Schnakenburg and P. West, {\sl Kac-Moody Symmetries of
IIB supergravity}, Phys. Lett. {\bf B 517} (2001) 137-145, {\tt
hep-th/0107181} 
\item{[11]} E. Bergshoeff, Mess de Roo, S. Kerstan and F. Riccioni, {\it
IIB Supergravity Revisited}, hep-th/0506013. 
\item{[12]} A. Kleinschmidt, I. Schnakenburg and P. West, {\sl
Very-extended Kac-Moody algebras and their interpretation at low
levels}, {\tt hep-th/0309198}
\item{[13]} P. West, {\it  The  IIA, IIB and eleven dimensional 
theories and their common $E_{11}$ origin}, hep-th/0402140. 
\item{[14]}  T. Damour, M. Henneaux and H. Nicolai, {\sl $E_{10}$ and a
``small tension expansion'' of M-theory}, Phys. Rev. Lett. {\bf 89}
(2002) 221601, {\tt hep-th/0207267}
\item{[15]} F. Englert, L. Houart, {\it 
    ${\cal G}^{+++}$ invariant formulation of gravity and M-theories: 
    Exact BPS solutions}, {\tt hep-th/0311255} 
\item{[16]}  M. R. Gaberdiel, D. I. Olive and P. West, {\sl A
class of Lorentzian Kac-Moody algebras}, Nucl. Phys. {\bf B 645}
(2002) 403-437,  hep-th/0205068. 
\item{[17]} P. West, {\it Very Extended $E_8$ and $A_8$ at low levels,
Gravity and Supergravity}, Class.Quant.Grav. 20 (2003) 2393-2406,
hep-th/0212291.
\item {[18]} P. Messen and T. Ortin, {\it An SL(2,Z) multiplet of nine
-dimensional type II supergravity theories}, Nucl. Phys. B {\bf 541}
(1999) 195, hep-th/9806120; 
G. Dall'Agata, K. Lechner and M. Tonin, {\it D=10, N=IIB supergravity;
Lorentz-invariant actions and duality}, JHEP {\bf 9807} (1998) 017,
hep-th/9806140;
E. Bergshoeff, U. Gran and D. Roest, {\it Type IIB seven-brane solutions
from nine-dimensional domain walls},  
  Class. Quant. Grav. {\bf 19} (2002) 4207, hep-th/0203202.
\item {[19]} A. Borisov and V. Ogievetsky, 
Theory of dynamical affine and conformal 
symmetries as the theory of the gravitational field, 
Teor. Mat. Fiz. 21 (1974) 329. 
\item{[20]}  A. Kleinschmidt and H. Nicolai, {\it IIB supergravity and
E(10)}, Phys. Lett B {\it 606} (2005) 391, hep-th/0309198. 
\item{[21]} J. Polchinski, Phys. Rev. Lett. 75 (1995) 184. 
\item{[22]} P. West, {\it Brane dynamics, central charges and
$E_{11}$}, hep-th/0412336. 
\item{[23]} E. Bergshoeff, Mess de Roo, B. Janssen and T. Ortin, {\it
The super D9-brane and its truncations}, Nulc. Phys. {\bf B550} (1999)
289, hep-th/9901055. 
\item{[24]} F. Riccioni, {\it Space-filling branes in ten and nine
diemnsions}, Phys. lett. {\bf B711} (2005) 231, hep-th/0410185. 
\item{[25]} private communication from  Fabio Riccioni to appear in a
forthcoming paper by E. Bergshoeff, Mess de Roo, S. Kerstan, T. Ortin and
F. Riccioni.  

\end